\documentclass[aps,floatfix,pra,reprint,superscriptaddress]{revtex4-2}

\usepackage{amsmath}
\usepackage{amssymb}
\usepackage{amsthm}
\usepackage{bbm}
\usepackage{color}
\usepackage{float}
\usepackage[T1]{fontenc}
\usepackage{graphicx}
\usepackage{hyperref}
\usepackage[utf8]{inputenc}
\usepackage{mathtools}
\usepackage{orcidlink}
\usepackage{physics}
\usepackage{times}
\usepackage{txfonts}
\usepackage{soul}
\usepackage{xcolor}

\hypersetup{
    colorlinks=true,linkcolor=blue,citecolor=blue,
    filecolor=blue,urlcolor=blue,breaklinks=true
}

\newcommand{\orcid}[1]{\href{https://orcid.org/#1}
{\includegraphics[width=7pt]{orcid.png}}}

\theoremstyle{definition}

\def\be{\begin{equation}}
\def\ee{\end{equation}}

\def\bc{\begin{center}}
\def\ec{\end{center}}
\def\bal{\begin{align}}
\def\eal{\end{align}}

\newcommand{\defis}{
  Departamento de Física,
  Universidade Estadual de Ponta Grossa
  84030-900 Ponta Grossa, Paraná, Brazil
}

\newcommand{\demat}{
  Departamento de Matemática e Estatística,
  Universidade Estadual de Ponta Grossa,
  84030-900 Ponta Grossa, Paraná, Brazil
}
\newcommand{\qpqi}{
  QPQI Group,
  Universidade Estadual de Ponta Grossa,
  84030-900 Ponta Grossa, Paraná, Brazil
}

\begin{document}

\title{
Continuous limit of the square well problem in quantum mechanics
}

\author{Matheus D. Moro\orcidlink{0009-0009-8709-2414}}
\email{matheusdinizmr01@gmail.com}
\affiliation{\qpqi}

\author{Thiago T. Tsutsui\orcidlink{0009-0001-1654-0330}}
\email{takajitsutsui@gmail.com}
\affiliation{\qpqi}

\author{Antonio S. M. de Castro\orcidlink{0000-0002-1521-9342}}
\email{asmcastro@uepg.br}
\affiliation{\qpqi}
\affiliation{\defis}

\author{Fabiano M. Andrade\orcidlink{0000-0001-5383-6168}}
\email{fmandrade@uepg.br}
\affiliation{\qpqi}
\affiliation{\demat}

\date{\today}

\begin{abstract}
The free-particle and square-well potentials are two of the most emblematic problems in quantum mechanics, illustrating essential concepts such as matter waves, energy quantization, and bound states. It is therefore natural to consider how the free-particle solutions emerge from the square well as the width approaches infinity. In this work, we present a systematic procedure to demonstrate this transition by applying a Fourier transform to the wave equation.
\end{abstract}

\maketitle

\section{Introduction}
\label{sec:introduction}
The Schr\"{o}dinger's equation stands as one of the most emblematic equations in physics, describing the time evolution of the wave function of a non-relativistic quantum particle \cite{Martins2025}, and reads
\begin{equation}
    \label{eq:totalscho}
    i\hbar\frac{\partial}{\partial t}\psi(\mathbf{r},t) = -\frac{\hbar^2}{2m}\nabla^2\psi(\mathbf{r},t)+U(\mathbf{r},t)\psi(\mathbf{r},t),
\end{equation}
where $\hbar=h/2\pi$ is the reduced Planck constant, $m$ is the non-relativistic particle mass, and $U(\mathbf{r},t)$ is the potential energy, which may be time-dependent. 
The wavefunction is a complex function of a real variable, denoted by $\psi(\mathbf{r},t)$.

The year 2026 marks the centenary of the papers in which Erwin Schrödinger first presented this equation, introducing wave mechanics to quantum theory \cite{SCHRODINGER1926A,SCHRODINGER1926B,SCHRODINGER1926C,SCHRODINGER1926D}.
This framework emerged as an alternative to matrix mechanics, which Werner Heisenberg had introduced a year earlier \cite{HEISENBERG1925}.
Both approaches proved instrumental in consolidating quantum mechanics as a theory, and the framework's subsequent success is linked to the contributions of both Schr\"{o}dinger and Heisenberg.
The theory, forged through the contributions of Schrödinger and other equally important scientists, proved remarkably successful -- and in the decades that followed, its non-classical aspects, such as superposition and entanglement, gave rise to the rapidly developing fields of quantum information and computation \cite{Nielsen2010}.
Moreover, quantum theory has also found applications in fields as diverse as biology \cite{Cirino2025} and economics \cite{DeBacker2025,Scursulim2026}.
From a technological standpoint, the theory enabled numerous innovations \cite{Singh1996}, including transistors \cite{ROSS1998,TIPLER2012} and lasers \cite{Schawlow1958,GERRY2005}.

Although the modern approach to quantum mechanics encompasses both wave and matrix mechanics \cite{SAKURAI2020}, wave mechanics remains particularly important in undergraduate courses \cite{TIPLER2012,EISBERG1985}.
In this context, two commonly studied problems are the free particle and the square well \cite{Lin2010}.
The former exhibits a continuous energy spectrum and plane-wave solutions that extend throughout all space; the latter introduces bound states and energy quantization, with the wave function confined within the well.

In this work, we examine the derivation of free particle wave functions from the square well solutions, a procedure that entails mathematical subtleties that go beyond simply taking the well's length to infinity. 
Besides being an instructive mathematical exercise, this derivation also holds physical interest, as it reveals the transition from discrete to continuum spectrum.
Our purpose is primarily didactic and concerns two problems of significant pedagogical value \cite{Belloni2014,GRIFFITHS2018}.

In our derivation, we employ the momentum-space representation of the wave function. 
While this representation has been used previously, for instance, to analyze boundary discontinuities \cite{Cummings1977} or to find momentum-space eigenfunctions for the infinite square well \cite{Rojo_2020}, our focus is different. 
Instead, we examine the transition from discrete to continuous energy eigenvalues and analyze at how the well's width dictates the separation between these levels.

This paper is organized as follows.
We begin in Sec. \ref{sec:free_particle} by reviewing the known solution for the free particle.
Building on this foundation, Secs. \ref{sec:fsquare_well} and \ref{sec:insquare_well} analyze the finite and infinite square well potentials, respectively.
Finally, in Sec. \ref{sec:discon}, we demonstrate how the free-particle continuum solutions emerge as the natural limit of a finite square well when its width tends to infinity by taking an appropriate Fourier Transform.
A summary of our findings is presented in Sec. \ref{sec:summ}.

\section{Free Particle}
\label{sec:free_particle}
Schr\"{o}dinger's seminal 1926 collection of papers, \emph{Quantisation as an Eigenvalue Problem} \cite{SCHRODINGER1926A,SCHRODINGER1926B,SCHRODINGER1926C,SCHRODINGER1926D}, introduced wave mechanics.
This formulation of microscopic phenomena was remarkably successful, yielding, among its achievements, the spectrum of the hydrogen atom.
Moreover, for many physicists of the time, it presented a more intuitive alternative to the matrix mechanics of Heisenberg, Born, and Jordan \cite{BELLER1983,HEISENBERG1925,jammer1989conceptual}.
However, some fundamental problems, such as the free electron, remained unresolved.
The reason was that the wave groups expanded without limit, which is incompatible with localized energy-exchange phenomena \cite{Martins2025}.
Here, for simplicity, we analyze the one-dimensional system.
In this case, the potential energy is null, $U(x,t)=0$, resulting in the ``free'' Schrödinger equation,
\begin{equation}
    \label{eq:free_schro}
    i\hbar\frac{\partial}{\partial t}\psi_{\text{free}}(x,t)=-\frac{\hbar^2}{2m}\frac{\partial^2}{\partial x^2}\psi_{\text{free}}(x,t).
\end{equation}

By the method of separation of variables \cite{Arfken2013}, the solution of Eq. \eqref{eq:free_schro} is expressed as the product of a spatial function and a time function,
\begin{equation}
    \label{eq:separated}
    \psi_{\text{free}}(x,t)=\chi_{\text{free}}(x)\xi_{\text{free}}(t).
\end{equation}
The corresponding two parts take the forms
\begin{subequations}
\begin{align}
\xi_{\text{free}}(t) & =e^{-iEt/\hbar}, \label{eq:time_sol_free} \\
\chi_{\text{free}}(x) &=Ae^{ip_xx/\hbar},\label{eq:spat_sol_free}
\end{align}
\end{subequations}
with $A$ as a normalization constant. 
Consequently, $\psi_{\text{free}}(x,t)$ can be written as
\begin{equation}
    \label{eq:total_sol}
    \psi_{\text{free}}(x,t)=Ae^{i(p_xx-Et)/\hbar}.
\end{equation}
However, Eq. \eqref{eq:total_sol} is not normalizable, since the integral over all space implies that $A$ tends to infinity. 
To address this issue, we employ wave packets. 
We assume that our solution can be written as the superposition of plane waves: 
\begin{equation}
    \label{eq:free}
    \psi_{\text{free}}(x,t)=\frac{1}{\sqrt{2\pi\hbar}}\int_{\text{all $p_x$ space}}\phi_{\text{free}}(p_x)e^{i(p_xx-Et)/\hbar}dp_x.
\end{equation}
The above equation is obtained using the Fourier Inverse Transform \cite{Arfken2013}, relating the spatial function to a superposition of momentum eigenfunctions. 
One of the main features of the free particle case is the continuous energy spectrum \cite{GRIFFITHS2018}.

\section{Finite Square Potential Well}
\label{sec:fsquare_well}
When Schr\"{o}dinger presented the equation that bears his name, he was mainly concerned with the practical problems of his time. 
This is why the first problems to which he applied his equations were the Hydrogen atom, the harmonic oscillator, the rigid rotor, and the Stark effect \cite{Belloni2014}.
Thus, the square well problem, which examines a particle trapped in a region of zero potential energy, 
was not among the selection of problems analyzed by Schr\"{o}dinger.
The first appearance of the infinite square well potential was probably in 1930, in Mott's work \cite{Gamow1966,Belloni2014}.
Further pedagogical and historical details can be found in the literature \cite{Belloni2014}.

As in the previous section, we can again employ the separation of variables method.
The temporal solution remains the same, since only the potential has changed. 
In this context, the time-independent Schrödinger equation is 
\begin{equation}
    \label{eq:squarew}
    -\frac{\hbar^2}{2m}\frac{d^2}{dx^2}\psi(x)+U(x)\psi(x)=E\psi(x),
\end{equation}
in which $U(x)$ is the potential and $E$ is the total energy.
In the context of the finite square well potential of length $L$, the potential energy $U(x)$ is expressed as follows 
\begin{equation}
    \label{eq:potential}
    U(x)=\left\{
    \begin{array}{l} U_{0}, \quad  x \leq-L/2, \\
    \vspace{-8pt} \\
   \,\, 0, \quad -L/2 < x < L/2, \\
    \vspace{-8pt} \\
    U_{0}, \quad -x \geq L/2. 
    \end{array}\right .
\end{equation}

For each region, we have a value of $U(x)$ associated with a different wave function. 
Therefore, we have three Schrödinger equations to solve: 
\begin{subequations}
    \begin{align}
    &-\frac{\hbar^2}{2m}\frac{d^2}{dx^2}\psi_{1}(x)+U_{0}\psi_{1}(x)=E\psi_{1}(x), \label{eq:squarewa} \\ 
    &-\frac{\hbar^2}{2m}\frac{d^2}{dx^2}\psi_{2}(x)=E\psi_{2}(x)                   \label{eq:squarewb}, \\
    &-\frac{\hbar^2}{2m}\frac{d^2}{dx^2}\psi_{3}(x)+U_{0}\psi_{3}(x)=E\psi_{3}(x)  \label{eq:squarewc}. 
\end{align}

\begin{figure*}
    \label{fig:finite}
    \centering
    \includegraphics[width=0.75\linewidth]{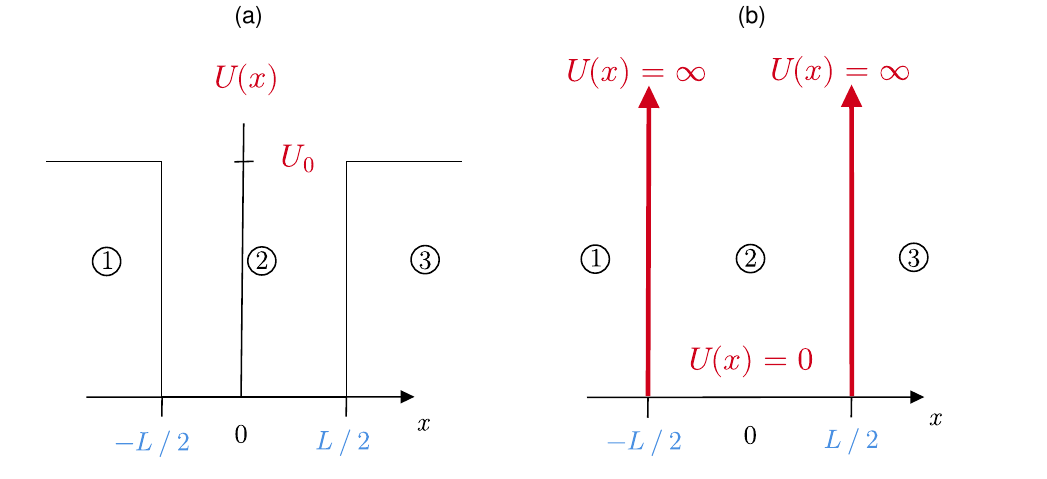}
    \caption{Schematic representation of the (a) finite and (b) infinite square well.}
\end{figure*}

\end{subequations}
Equation \eqref{eq:squarewb} is also associated with the Schrödinger equation for the infinite square well potential. For this reason, the calculation of its eigenfunctions and eigenvalues is deferred to the next section.
The solutions to Eqs. \eqref{eq:squarewa} and \eqref{eq:squarewc} have the same general form but possess different boundary conditions. Given that, we will solve just one of the equations and later apply the appropriate boundary conditions. Consider, then, the following:
\begin{equation}
    \label{eq:squarewj}
    -\frac{\hbar^2}{2m}\frac{d^2}{dx^2}\psi_{j}(x)+U_{0}\psi_{j}(x)=E\psi_{j}(x),\quad j=1,3.
\end{equation}
With elementary algebra, we rearrange the differential equation:
\begin{align}
    \label{eq:squarew_rearrenged}
    &\frac{d^2}{dx^2}\psi_{j}(x)+\kappa^2\psi_{j}(x)=0,
\end{align}
with $\kappa = \sqrt{ 2m(U_{0}-E)/\hbar^2 }$.
We consider an ansatz solution and its second-order derivative in the form of
\begin{subequations}
    \begin{align}
    \psi_{j}(x) &= e^{\lambda x}, \label{eq:ansatz} \\
    \frac{d^2}{dx^2}\psi_{j}(x) &= \lambda^2 e^{\lambda x}. \label{eq:derivative_ansatz}
\end{align}
\end{subequations}
Substituting the above equations into  Eq. \eqref{eq:squarew_rearrenged}, we obtain the roots of the corresponding characteristic polynomials \cite{boyce2017elementary} in the form
\begin{align}
    \label{eq:lambda_alpha}
    &\lambda= \pm \kappa.
\end{align}
Clearly, the difference between $E$ and $U_{0}$ results in different solutions for the wave function. 
Here we assume $U_{0} \geq E$, so that $\kappa$ is a real number and the wave function is associated with a bound state.
Consequently, the general solution is given by 
\begin{equation}
    \label{eq:gen_sol}
    \psi_{j}(x)=A_{j}e^{\kappa x}+B_{j}e^{-\kappa x}.
\end{equation}
Now, we apply the boundary conditions. 
In the leftmost region ($j=1$), we set $B_1 = 0$, as the term $e^{-\kappa x}$ would diverge as $x \to -\infty$, implying a greater probability of finding the particle as $x \rightarrow -\infty$.
Similarly, in the rightmost region ($j=3$), we set $A_3 = 0$, since $e^{\kappa x}$ diverges as $x \to \infty$.
Accordingly, outside the potential well, the solutions take the form:
\begin{subequations}
\begin{align}
    \psi_{1}(x)= {} & A_{1}e^{ \kappa x}, \qquad x \leq  -L/2, \label{eq:sol1} \\
    \psi_{3}(x)= {} & B_{3}e^{-\kappa x}, \qquad  x \geq   L/2. \label{eq:sol3}
\end{align}
\end{subequations}

\section{Infinite Square Well}
\label{sec:insquare_well}

To obtain a solution for the infinite square well potential, we impose the following prescription for the potential energy:
\begin{equation}
    \label{eq:infinte_potential}
    U(x)=\left\{
    \begin{array}{l} 
    0, \quad -L/2<x < L/2, \\ 
     \vspace{-8pt} \\
    \infty, \quad -x \geq L/2\, \text{and}\, x\leq L/2. 
    \end{array}\right .
\end{equation}
We note that in this case, the wave number $\kappa$ diverges,
\begin{align}
    \label{eq:potential_limit}
    \lim_{ U_{0} \to \infty }\kappa =\lim_{ U_{0} \to \infty }\sqrt{ \frac{2m}{\hbar^2}(U_{0}-E) }.
\end{align}
Consequently, the wave function in regions 1 and 3 must vanish
\begin{align}
    \label{eq:sol_limit}
    \lim_{ U_{0} \to \infty }\psi_{1}(x)=\lim_{ U_{0} \to \infty }\psi_{3}(x)=0.
\end{align}
Next, we solve Eq.  \eqref{eq:squarewb} in region 2:
\begin{align}
    \label{eq:squarewb2}
    \frac{d^2}{dx^2}\psi_{2}(x)&+k^2\psi_{2}(x)=0,
\end{align}
with $k=\sqrt{2mE/\hbar^2}$.
Using the same reasoning as used in the previous section, we suppose a solution of the form $\psi_{2}(x)=e^{\gamma x}$, leading to
\begin{subequations}
    \begin{align}
    \frac{d^2}{dx^2}\psi_{2}(x)&=\gamma^2e^{\gamma x}. \label{eq:derivative_ansatz2}
\end{align}
\end{subequations}
In this manner, we obtain the following
\begin{equation}
  \label{eq:gamma_beta}
    \gamma = \pm i k.   
\end{equation}
and the wave function in region 2 is given by
\begin{equation}
    \label{eq:sol2}
    \psi_{2}(x)=A_{2}e^{i k x}+B_{2}e^{-i k x}.
\end{equation}
Using Euler's identity, $ e^{\pm i k x}=\cos(k x)\pm i\sin(k x)$, we can rewrite our solution as follows 
\begin{align}
    \label{eq:sol1_rearenged}
\psi_{2}(x) = {} & C_{2}\cos(k x)+D_{2}\sin( kx).
\end{align}
According to the boundary conditions, the wave function must vanish when $x=-L/2$ and $x=L/2$. Applying those to Eq. \eqref{eq:sol1_rearenged}, we have the following
\begin{subequations}
    \begin{align}
    \label{eq:bound1}
    C_{2}\cos(kL/2) - D_{2}\sin(kL/2) = {} & 0,
    \\
    \label{eq:bound2}
    C_{2}\cos(kL/2) + D_{2}\sin(kL/2) = {} & 0.
\end{align}
\end{subequations}
Combining Eqs. \eqref{eq:bound1} and \eqref{eq:bound2}, it results in two families of solutions that are distinguished by their parity properties:
\begin{align}
    \label{eq. 29} 2C_{2}\cos(k L/2)&=0, \\
    \label{eq.30} 2D_{2}\sin(k L/2)&=0.
\end{align}
We denote the odd and even families of solutions as $\xi(x)$ and $\phi(x)$, and simplify the notation by setting  $C_{2} = A$ and $D_{2} = B$, which leads to  solutions of the form:
\begin{subequations}
    \begin{align}
    \label{eq:fam1}\psi_n^{\rm even}(x) = {} & A\cos(kx),\quad \text{ for } \cos(kL/2)=0, \\
    \label{eq:fam2}\psi_n^{\rm odd}(x) = {} & B\sin(kx), \quad \text{ for }\sin(kL/2)=0.
\end{align}
\end{subequations}
Thus, applying the boundary conditions results in a transcendental equation for $k$. 
In this case, the solutions are found by inverting the trigonometric functions, leading to:
\begin{subequations}  
\label{eq:trans}
\begin{align}
\label{eq:transcos2}
 k_{n}=\frac{n\pi}{L},\quad \text{ for } n=1,3,5,\dots \\
\label{eq:trans2}
 k_{n}=\frac{n\pi}{L},\quad \text{ for } n=2,4,6,\dots 
\end{align}
\end{subequations}
Thus, we obtain an unnormalized wave function for region 2 as follows
\begin{equation}
    \label{eq:non_normalized_solution}
    \Phi_{n}(x)=\left\{\begin{array}{l} 
    \psi_{n}^{\rm even}(x)=A \cos(k_{n}x), \quad \space \displaystyle k_{n}=\frac{n\pi}{L},\space n=1,3,5,\dots, \\
                    \\
    \psi_{n}^{\rm odd}(x)=B \sin(k_{n}x), \quad  \space \displaystyle k_{n}=\frac{n\pi}{L},\space n=2,3,6,\dots. \end{array}\right.
\end{equation}

The case $n=0$ is ignored, as it corresponds to the trivial solution representing the absence of a particle in the well. 
The above solutions are quantum analogs of the classical problem of standing waves on a string with fixed ends. 
Although its first appearance was in the Mott textbook \cite{Mott1930}, this conceptual framework dates back to at least 1900, with Jeans' box or Jeans' cube: a cube with sides consisting of perfectly reflecting material \cite{Belloni2014}.

The normalization constants $A$ and $B$ can be obtained from the normalization condition:
\begin{subequations}
    \begin{align}
    \label{eq:norm1}
    \int^{L/2}_{-L/2}|\psi_{2}^{\rm even}(x)|^2dx&=1, \\
    \label{eq:norm2}\int^{L/2}_{-L/2}|\psi_{2}^{\rm odd}(x)|^2dx&=1.
\end{align}
\end{subequations}
Thus, from Eqs. \eqref{eq:non_normalized_solution} and \eqref{eq:norm1}, and by changing the integration variable to $u=n\pi x/L$, we have the following 
\begin{align}
    \label{eq:int1}
    \int_{-n\pi/2}^{n\pi/2} \cos^2(u)du=\frac{n\pi}{A^2 L},
\end{align}
and using the formula:
\begin{equation}
    \label{eq:indentity1}
    \int^b_{a}\cos^n(u)du=
    \left. \frac{\cos^{n-1}(u)\sin(u)}{n}\right|_a^b+\frac{n-1}{n}\int_{a}^b\cos^{n-2}(u)du,
\end{equation}
we obtain the normalization constant $A$, which is independent of $n$, as 
\begin{align}
    \label{eq:cons_normalization1}
    A=\sqrt{ \frac{2}{L} }.
\end{align}
Therefore, our even normalized wavefunction reads
\begin{equation}
    \label{eq:norm_wavefunction1}
    \psi_{n}^{\rm even}(x)=\sqrt{ \frac{2}{L} }\cos(k_{n}x).
\end{equation}
Using the same procedure, but emplyong the formulae
\begin{equation}
    \label{eq:indentity2}
    \int^b_{a}\sin^n(u)du=
    \left. -\frac{\cos(u)\sin^{n-1}(u)}{n}\right|_a^b+\frac{n-1}{n}\int_{a}^b\sin^{n-2}(u)du.
\end{equation}
we obtain
\begin{align}
    \label{eq:cons_normalization2}
    B=\sqrt{ \frac{2}{L} }.
\end{align}
Accordingly, the odd family of solutions is given by:
\begin{equation}
    \label{eq:norm_wavefunction2}
    \psi_{n}^{\rm odd}(x)=\sqrt{ \frac{2}{L} }\sin(k_{n}x).
\end{equation}
Thus, the full solution is as follows 
\begin{equation}
    \label{eq:normalized_eigenfunction}
    \Phi_{n}(x)=\left\{\begin{array}{l} 
    \displaystyle\psi_{n}^{\rm even}(x)=\sqrt{ \frac{2}{L} }\cos(k_{n}x), \quad k_{n}=\frac{n\pi}{L},\space n=1,3,5,\dots, \\
    \displaystyle\psi_{n}^{\rm odd}(x)=\sqrt{ \frac{2}{L} }\sin(k_{n}x), \quad k_{n}=\frac{n\pi}{L},\space n=2,4,6,\dots. \end{array}\right.,
\end{equation}
and the eigenenergies are
\begin{align}
    \label{eq:eigenergies}
    E_{n}&=\frac{n^2\hbar^2 \pi^2}{2mL^2},\quad \textrm{with}\ n=1,2,3,\dots 
\end{align}

\section{From discrete to continuous spectrum}
\label{sec:discon}

In the limit $L \rightarrow \infty$, the potential becomes infinitely wide. 
As the width of the well increases, the spacing between consecutive values of $k_n$ decreases, and the discrete spectrum approaches a continuum. 
In this limit, it becomes natural to describe the state in terms of a continuous momentum distribution.

\begin{figure*}
    \centering
    \includegraphics[width=0.75\linewidth]{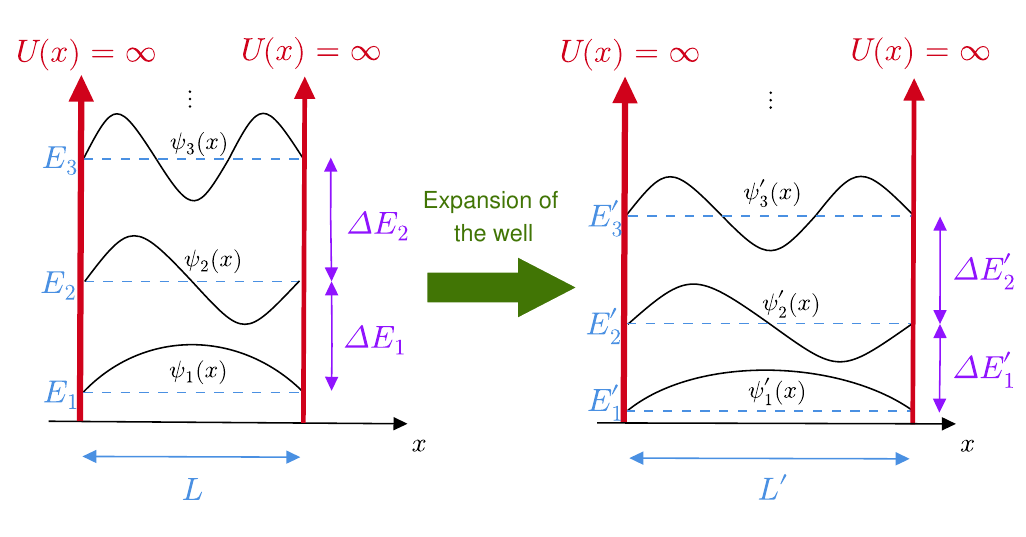}
    \caption{Schematic and out-of-scale illustration of the expansion of the well and its effects on the energy difference between levels. 
    The expansion $\left(L^{\prime}>L\right)$, results in $E_n^{\prime}<E_n$ and $\Delta E_n^{\prime}<\Delta E_n$.}
    \label{fig:energy}
\end{figure*}

Firstly, we analyze how the discrete energy spectrum becomes a continuum.
So, consider the energy for the levels $n$ and $n+1$, and the difference between these two adjacent energy levels is given by 
\begin{align}
    \label{eq:energy_difference}
    \Delta E = {} & E_{n+1}-E_{n},\notag \\
    = {} & (2n+1)\frac{\hbar^2\pi^2}{2mL^2}.
\end{align}
Taking the limit as $L \rightarrow \infty$, we obtain $\lim_{ L \to \infty } \Delta E = 0$, indicating that as the well length approaches infinity, a transition occurs from a discrete energy spectrum to a continuum spectrum.
This result is consistent with the behavior of a free particle, where the energy is no longer restricted to quantized levels.
This process is schematically illustrated in Fig. \ref{fig:energy}.

Next, we examine the behavior of the wavefunction in this limit.
Consider a general state written as a superposition of the eigenstates of the finite square well as follows 
\begin{align}
    \label{eq:superposition_rearenged}
    \psi(x)= N\sum_{n=1}^{\infty}\left[a_{n}\sin(k_{n}x)+b_{n}\cos(k_{n}x)\right], 
\end{align}
where $a_n$ and $b_n$ are determined by
\begin{subequations}
    \begin{align}
    \label{eq:fourier_coefficient1}a_{n}&=\frac{2}{L}\int_{-L/2}^{L/2}\psi(x)\sin\left( \frac{n\pi x}{L} \right)dx,\quad \text{ even } n, \\
    \label{eq:fourier_coefficient2}b_{n}&=\frac{2}{L}\int_{-L/2}^{L/2}\psi(x)\cos\left( \frac{n\pi x}{L} \right)dx,\quad \text{ odd } n,
\end{align}
\end{subequations}
and $N$ is a proper normalization constant left unspecified at this stage because the normalization of the wave function changes when passing from the discrete to the continuum.
Using  Euler's identity, it follows that
\begin{align}
    \label{eq:superposition_complex}
    \psi(x) = {} & N \sum_{n=1}^{\infty}\left[a_{n}\sin(k_{n}x)+b_{n}\cos(k_{n}x)\right], \notag\\
    = {} & N \sum_{n=1}^{\infty}\left[a_{n}\left(\frac{e^{ik_{n}x}-e^{-ik_{n}x}}{2i}\right)+b_{n}\left(\frac{e^{ik_{n}x}+e^{-ik_{n}x}}{2}\right)\right],\notag \\
    = {} & N \sum_{n=1}^{\infty}\left[\left(\frac{b_{n}-ia_{n}}{2}\right)e^{ik_{n}x}+\left(\frac{b_{n}+ia_{n}}{2}\right)e^{-ik_{n}x}\right],\notag \\
   = {} & N\left[ \sum^{\infty}_{n=1}\left(\frac{b_{n}-ia_{n}}{2}\right)e^{in\pi x/L}+\sum^{-1}_{n'=-\infty}\left(\frac{b_{-n'}+ia_{-n'}}{2}\right)e^{in'\pi x/L}\right],
\end{align}
where we introduced the dummy index $n' = -n$.
If we switch from $n$ to $-n$ in Eqs. \eqref{eq:fourier_coefficient1} and \eqref{eq:fourier_coefficient2}, we obtain:
\begin{subequations}
    \begin{align}
    \label{eq:fourier_coefficient11}a_{-n}&=\frac{2}{L}\int_{-L/2}^{L/2}\psi(x)\sin\left( \frac{-n\pi x}{L} \right)dx, & \Rightarrow & & a_{-n}= {} & -a_{n},
    \\
    \label{eq:fourier_coefficient22}b_{-n}&=\frac{2}{L}\int_{-L/2}^{L/2}\psi(x)\cos\left( \frac{-n\pi x}{L} \right)dx,  & \Rightarrow & & b_{-n}= {} & b_{n}.
\end{align}
\end{subequations}
Then, using the relations above, Eq. \eqref{eq:superposition_complex} becomes
\begin{align}
    \label{eq:complex_form}
    \psi(x)= {} & N\sum^{\infty}_{n=1}\left(\frac{b_{n}-ia_{n}}{2}\right)e^{in\pi x/L}+\sum^{-1}_{n=-\infty}\left(\frac{b_{-n}+ia_{-n}}{2}\right)e^{in\pi x/L},\notag \\
    = {} & N\sum^{\infty}_{n=1}\left(\frac{b_{n}-ia_{n}}{2}\right)e^{in\pi x/L}+\sum^{-1}_{n=-\infty}\left(\frac{b_{n}-ia_{n}}{2}\right)e^{in\pi x/L}, \notag\\
    = {} & N\sum_{n=-\infty}^{\infty}c_{n}e^{in\pi x/L},\
\end{align}
with $c_{n}$ given by 
\begin{align}
    \label{eq:fouerir_complex_coefficient}
    c_{n} = {} & \frac{b_{n}-ia_{n}}{2} \nonumber \\
    = {} &  
    \frac{1}{L}\int_{-L/2}^{L/2}\psi(x)\left[\cos\left( \frac{n\pi x}{L} \right)-i\sin\left( \frac{n\pi x}{L} \right)\right]dx,\notag \\
    = {} & \frac{1}{L}\int_{-L/2}^{L/2}\psi(x)e^{-in\pi x/L}dx.
\end{align}
The limit $L\to \infty$ requires some care because the direct calculation results in $c_{n}=0$. 
However, we know that $k_{n}=n \pi/L$ and, therefore, $\Delta k_{n}=(\pi/L)\Delta n$. 
The distance between two adjacent energy levels is $\Delta n=1$, which allows us to write:
\begin{align}
    \label{eq:equality_to_1}
    1=\frac{L\Delta k_{n}}{\pi}.
\end{align}
Thus, inserting this quantity into Eq. \eqref{eq:complex_form}, we find 
\begin{align}
    \label{eq:new_complex_form}
    \psi(x)= {} & N\sum_{n=-\infty}^{\infty}c_{n}e^{ik_n x}\frac{L\Delta k_{n}}{\pi},\notag \\
    = {} & N\sum^{\infty}_{n=-\infty}\frac{c_{n}L}{\pi}e^{i k_nx}\Delta k_{n}.
\end{align}
This motivates the introduction of a new coefficient $\phi(k_n)$ given by
\begin{align}
    \label{eq:new_coefficient}
 \phi(k_n) = {} & \frac{L}{\pi}c_{n} \notag \\
           = {} & \frac{1}{\pi}\int_{-L/2}^{L/2}\psi(x)e^{-i k_n x}dx, 
\end{align}
in such a way that Eq. \eqref{eq:new_complex_form} becomes
\begin{equation}
    \label{eq:new_coefficient_complex_form}
    \psi(x)= N \sum^{\infty}_{\frac{Lk_{n}}{\pi}=-\infty}\phi(k_n)e^{i k_n x}\Delta k_{n}.
\end{equation}
Taking the limit $L\to \infty$, it implies that $\Delta k_n \to dk $ and $k_n\to k$. 
Moreover, in this limit, following Ref. \cite{Amaku2020}, the normalization of the wave function of continuum states leads to the choice $N=1/2$. 
Thus, we obtain
\begin{equation}
\label{eq:psi_Ck}
    \psi(x) = \frac{1}{2}\int^{\infty}_{-\infty}\phi(k)e^{ikx}dk.
\end{equation}
and
\begin{align}
    \label{eq:int_complex_form1}
    \phi(k)= {} & \lim_{ L \to \infty }\phi(k_n) \nonumber \\
    = {} & \frac{1}{\pi}\int_{-\infty}^{\infty}\psi(x)e^{-ikx}dx,
\end{align}
The equations above are similar to a Fourier Transform \cite{SAKURAI2020}.
Consequently, to obtain the free particle solution from the infinite square well potential, we follow a procedure analogous to the transition from a Fourier Series to a Fourier Transform. 
This approach, along with some analogous derivations, can be found in the literature \cite{butkov1968mathematical,Cummings1977,Rojo_2020}.
Introducing the symmetric Fourier transform convention commonly used in quantum mechanics \cite{SAKURAI2020} and using $p=\hbar k$, Eqs. \eqref{eq:int_complex_form1} and \eqref{eq:psi_Ck} can be rewritten as
\begin{subequations}
    \begin{align}
    \label{eq:split_form1}
    \psi(x)&=\frac{1}{\sqrt{ 2 \pi \hbar }}\int^{\infty}_{-\infty}\phi(p)e^{ipx/\hbar}dp,
    \\
    \label{eq:split_form2}\phi(p)&=\frac{1}{\sqrt{2 \pi \hbar}}\int_{-\infty}^{\infty}\psi(x)e^{-ipx/\hbar}dx.
\end{align}
\end{subequations}
To recover the solution of the momentum eigenfunction of the free particle with a well-defined linear momentum $p'$, we consider the particular choice
\begin{equation}
    \label{eq:dirac_delta_p}
    \phi(p)=\delta(p-p').
\end{equation}
Thus, our wavefunction is obtained by substituting  Eq. \eqref{eq:dirac_delta_p} into Eq. \eqref{eq:split_form1},
\begin{align}
    \label{eq:free_particle_sol}
    \psi(x) = {} & \frac{1}{\sqrt{2 \pi \hbar}}\int^{\infty}_{-\infty}\delta(p-p')e^{ipx/\hbar}dp,\notag \\
   = {} &\frac{1}{\sqrt{2 \pi \hbar}}e^{ip'x/\hbar}.
\end{align}
The above wave function is just the familiar plane-wave solution for the free particle.

\section{Summary}
\label{sec:summ}

In this work, we have shown how to derive the free-particle wave function from the square well potential by taking the continuum limit via a Fourier transform.
We have also reviewed both problems separately to make this work self-contained.

Research in physics education indicates that students in quantum mechanics curricula who begin with spin systems before transitioning to continuous systems encounter difficulties when moving to problems in continuous space, such as the infinite square well \cite{Solorio2025}.
A study such as the present one, which deepens the understanding of this problem, may therefore prove useful.

\section*{Acknowledgments}
This work was partially financed by the Coordenação de Aperfeiçoamento
de Pessoal de Nível Superior (CAPES, Finance Code 001).
It was also supported by the Conselho Nacional de Desenvolvimento
Científico e Tecnológico (CNPq).
F.M.A. and A.S.M.C. acknowledge financial support from Fundação Araucária (Project No. 305)
F.M.A. acknowledges financial support by CNPq Grant No. 313124/2023-0.\\

\noindent\textbf{Conflict of interest}\\
The authors declare no conflicts of interest.\\

\noindent\textbf{Data availability}\\
No new data were created or analyzed in this study. This is purely a theoretical work, and all mathematical derivations and physical conclusions are explicitly detailed within the manuscript.

\bibliographystyle{apsrev4-2}
%

\end{document}